\documentclass[12pt,a4paper,onesided]{article}
\usepackage{amsmath, amssymb, amsthm}
\usepackage{authblk}
\usepackage[english]{babel}
\usepackage{bm}
\usepackage{color}
\usepackage{float}
\usepackage[T1]{fontenc}
\usepackage[hidelinks]{hyperref}
\usepackage[utf8]{inputenc}
\usepackage{multirow}
\usepackage{natbib}
\usepackage{sectsty}
\usepackage{setspace}
\usepackage{subcaption}
\usepackage{tabularx}
\usepackage{xfrac}
\usepackage{titlesec}
\usepackage[noline, boxed]{algorithm2e}
\usepackage{lscape}

\titlelabel{\thetitle.\quad}
\subsectionfont{\itshape}

\title{\textbf{A Change-Point Based Control Chart for Detecting Sparse Changes in High-Dimensional Heteroscedastic Data}}
\author[1,*]{Zezhong Wang}
\author[1]{Inez Maria Zwetsloot}
\affil{Department of Systems Engineering and Engineering Management, City University of Hong Kong, Tat Chee Avenue, Kowloon, Hong Kong}
\affil[*]{Corresponding author: Zezhong Wang, zezhowang3-c@my.cityu.edu.hk}

\begin{document}	
\maketitle

Total Words:4226
\newpage
\begin{center}
\textbf{A Change-Point Based Control Chart for Detecting Sparse Changes in High-Dimensional Heteroscedastic Data}
\end{center}

\noindent \textbf{Abstract}\\
Because of the curse-of-dimensionality, high-dimensional processes present challenges to traditional multivariate statistical process monitoring (SPM) techniques. In addition, the unknown underlying distribution and complicated dependency among variables such as heteroscedasticity increase uncertainty of estimated parameters, and decrease the effectiveness of control charts. In addition, the requirement of sufficient reference samples limits the application of traditional charts in high dimension low sample size scenarios (small $n$, large $p$). More difficulties appear in detecting and diagnosing abnormal behaviors that are caused by a small set of variables, i.e., sparse changes. In this article, we propose a change-point based control chart to detect sparse shifts in the mean vector of high-dimensional heteroscedastic processes. Our proposed method can start monitoring when the number of observations is a lot smaller than the dimensionality. The simulation results show its robustness to nonnormality and heteroscedasticity. A real data example is used to illustrate the effectiveness of the proposed control chart in high-dimensional applications. Supplementary material and code are provided online. \\[0.4cm]


   \nocite{*}    
   \newpage
\section{Introduction}\label{section 1}
In modern manufacturing systems, widely used sensor and internet technologies make it possible to collect hundreds of measurements related to the final product and its production over time. These measurements can be used to determine whether the process' performance is under control or whether corrective action is needed. Statistical process monitoring (SPM) tools for high-dimensional data are becoming popular for such applications. 

Detecting abnormality from high-dimensional data is a challenging task. Traditional multivariate control charts, such as the Hotelling's $T^2$ control chart and the MEWMA chart, are popular in monitoring multiple variables simultaneously. However, they are challenged by the ``curse-of-dimensionality'' which affects parameter estimates and signal detection ability significantly \citep{hastie2009elements}.

The accuracy of parameter estimates is an important factor of control charts' performances, but the uncertainty of estimates, especially of the estimated covariance matrix, grows rapidly with dimensionality \citep{zhang2020spatial}. In conventional monitoring schemes, large amounts of in-control (IC) data are needed for getting reliable parameter estimates before monitoring can start \citep{zamba2006multivariate}. One popular way to avoid directly estimating the convariance matrix is projecting the variables into a lower dimensional subspace by principal components analysis (PCA), and then monitoring the PCA scores. \citet{de2015overview} give an overview of PCA-based high-dimensional process monitoring techniques. 

PCA-based methods are generally designed under the assumption that the process follows a multivariate normal distribution, this assumption is violated in many high-dimensional processes applications \citep{zhang2020spatial}. The underlying distribution is usually unknown and challenging to verify. This situation inspires the development of nonparametric control charts, such as the spatial sign based method proposed by \citet{zou2011multivariate}. 

One drawback of the nonparametric method  is that it requires sufficient IC observations to estimate various parameters which affect the performance of control charts \citep{zou2011multivariate, zou2012spatial, holland2014control}. They can not be directly applied in high dimension low sample size (HDLSS) scenarios where the number of variables is much greater than the number of individual observations $(p>n)$, this is also referred to as short-run processes. Change-point based control charts have better performances in this scenario \citep{li2014self}. 

It is rare that all variables change simultaneously in a process, a small subset of variables are more likely to cause the abnormality. Under this sparsity assumption, the out-of-control (OC) signals can be easily buried in the noise of numerous variables \citep{shu2018distribution}. Sparsity also increases the difficulty to locate and identify assignable causes among variables \citep{wang2009high}. 

One overlooked character in SPM solutions for high-dimensional data is heteroscedasticity \citep{hong2018asymptotic}. A straightforward definition of heteroscedasticity is the inequality of error variance \citep{downs1979interpreting}. This character is deeply discussed in financial and economic fields, and is treated as a set of variances to be modeled \citep{engle2001sons}. It is rarely considered in manufacturing industries thought it would affect the accuracy of parameters estimate. Consequently, the performances of control charts are fallible. 

To overcome the three mentioned challenges of monitoring high-dimensional processes, 1) the HDLSS problem makes parameter estimation impossible, 2) sparsity of the change, and 3) heteroscedasticity in the data, we propose a change-point based control chart for detecting sparse mean shifts in high-dimensional heteroscedastic data. The change-point methods allow us to deal naturally with the first challenge. We apply a supremum-based test to deal with the second and third challenges. The proposed method can start monitoring when the sample size is much smaller than the dimension $(p\gg n)$. Another attraction is that this method is robust to both nonnormality and heteroscedasticity. We also propose a post-signal diagnosis method to estimate the change-point and assignable causes. Simulation and comparison results show that the proposed method has good performance in detecting large and sparse shifts from high-dimensional heteroscedastic data streams. Application of the methods is illustrated by a real-data example. 

The structure of this article is as follows. Section \ref{section 2} is a review of the related literature. In Section \ref{section 3}, we propose our novel change-point based control chart for detecting sparse shifts in the mean vector of possibly heteroscedastic non-normal data. Section \ref{section 4} shows the performance of the proposed method. In Section \ref{section 5}, we propose a post-signal diagnosis procedure. Section \ref{section 6} discusses a comparison with another method. In Section \ref{section 7}, a manufacturing process example is used to showcase the proposed method. Section \ref{section 8} concludes.

\section{Related Work}
\label{section 2}
Quite some research has been done to adapt classical SPM methods to high-dimensional monitoring problems. Combining dimensionality reduction techniques with control charts is one alternative to efficiently detect sparse changes in the high-dimensional setting. \citet{wang2009high}, \citet{zou2009multivariate}, \citet{zou2011lasso}, \citet{capizzi2011least}, \citet{jiang2012variable}, and \citet{abdella2017variable} applied different penalized likelihood-based variable selection algorithms, such as LASSO, to screen out suspicious variables that possibly make the process deviate from its normal stage. Next, conventional control charts are applied to the selected variables. 

These above discussed methods are efficient under the assumption of normal distributed data, which is usually violated. Nonparametric methods are developed to fill this gap. \citet{zou2015efficient} proposed an online monitoring scheme based on a goodness-of-fit (GOF) test for monitoring nonnormal and i.i.d. observations. \citet{shu2018distribution} proposed a distribution-free control chart by adapting a Minkowski distance based test.

The sparsity also increases the difficulty for signal diagnosis, one benefit of variable-selection algorithm is that the tasks of monitoring and diagnosis are naturally integrated and conveniently solved simultaneously \citep{jiang2012variable}. In \citet{zhang2020spatial}, they proposed a square-root LASSO based diagnosis framework. An effective tool to locate the occurrence time of changes in data sequences is change-point estimate, it has been widely used in univariate, multivariate, and profile monitoring problems. \citet{amiri2012change} reviewed the applications of change-point estimation techniques in process monitoring systematically. This type of method not only supports practitioners to identify assignable causes but also shows robustness to nonnormality. Therefore, change-point based methods are potentially useful in monitoring high-dimensional data streams. 

\citet{zou2011lasso} estimated the change-point in post-signal diagnosis phase. \citet{li2014self, huang2014high, chen2016distribution} also proposed change-point based control charts. The first one is developed from a two-sample location test proposed by \citet{chen2010two}, which is a sum-of-square type test.  It is more efficient in monitoring dense shifts. \citet{huang2014high} proposed a novel Reproducing Kernel Hilbert Space (RKHS)-based control chart which is robust to nonnormality and is able to detect a wide range of process changes include the sparse changes. The method proposed by \citet{chen2016distribution} is developed from \citet{bickel1969distribution}, and only dense scenarios are considered in the simulation part. One more attraction of the above change-point based schemes is that they are capable of starting monitoring with a small set of observations. 

The heteroscedasticity has been well studied in financial data. This type of data can be modeled by autoregressive conditional heteroscedastic (ARCH) models \citep{engle1982autoregressive} or generalized autoregressive conditional heteroskedastic (GARCH) models \citep{bollerslev1986generalized}. To monitor for shifts in heteroscedastic data, \citet{schipper2001control} and \citet{bodnar2009application} proposed control charts based on univariate and multivariate GARCH models respectively. Both methods used the same strategy that is monitoring the fitting errors between the target process and the observed process. It is challenging to apply the same strategy to high-dimensional scenarios, because of the computational complexity \citep{frisen2008financial}. Another alternative strategy for monitoring heteroscedastic data is using methods that are robust to inconsistent covariance matrices. 

The above literature offers solutions for either sparse changes in normal data or HDLSS methods for dense changes or method focused on heteroscedasticity. No method is able to do it all. Our objective is to develop a method to detect sparse changes in heteroscedastic and possibly non-normal data in the HDLSS scenarios. We will adapt the two-sample hypothesis tests of high-dimensional means proposed by \citet{chang2017simulation} to process monitoring in Section \ref{section 3}, because it allows for different covariance matrices between two samples. 

\section{Change-Point Based Control Chart}\label{section 3}
In this section we introduce our proposed method.
\subsection{Change-Point Based Monitoring Scheme}\label{3.1}
Given a series of independent and individual $p$-dimensional observations $\bm{X}_{i}=(X_{i1}, X_{i2}, ..., X_{ip})'$, $i=1, 2, ..., n$, which follow a multivariate normal distribution, the change-point model is
\begin{equation}\label{eq1}
H_{0}: \bm{X}_{i} \sim N_{p}(\bm{\mu}_{0}, \bm{\Sigma}_{0})
\hspace{20pt}\text{ and }\hspace{20pt}
H_{1}: \bm{X}_{i} \sim
\begin{cases} 
      N_{p}(\bm{\mu}_{0}, \bm{\Sigma}_{0}), & i\leq \tau \\
      N_{p}(\bm{\mu}_{1}, \bm{\Sigma}_{1}), & i>\tau,
   \end{cases}
\end{equation}
where $\bm{\mu}_{0}$, $\bm{\mu}_{1}$, $\bm{\Sigma}_{0}$, and $\bm{\Sigma}_{1}$ are the unknown IC and OC mean vectors and covariance matrices. The change-point $\tau$, which indicates that the process changes after $\bm{X}_{\tau}$, is also unknown and needs to be estimated. If only a mean shift occurs, $\bm{\Sigma}_{0}$ is equal to $\bm{\Sigma}_{1}$ in Equation (\ref{eq1}). 

Various high-dimensional hypothesis tests for the shifts in locations have been proposed, see, for example, \citet{randles2000simpler, zhang2002powerful, chen2010two, saha2017some}. These tests have been integrated into process monitoring tools that were reviewed in Sections \ref{section 1} and \ref{section 2}. High-dimensional test statistics can be categorized as sum-of-squares-based test statistic \citep{chen2010two}, and supremum-based test statistic \citep{randles2000simpler}. The former is powerful when there are many small differences between $\bm{\mu}_{0}$ and $\bm{\mu}_{1}$, in other words, the signals are dense but weak. However, this type of method could be ineffective under sparsity assumption, where the accumulation of all differences will not be greatly influenced by a few large differences especially in a big data stream. For the case that a few variables have large shifts, the signals are sparse but strong, a supremum-based test statistic should have better performance \citep{gregory2015two}. 

Our proposed control chart for detecting sparse mean shifts is based on the supremum-based-test statistics proposed by \citet{chang2017simulation}. \citet{chang2017simulation} proposed four tests. The first two are supremum-based test based on a non-studentized test statistics and a studentized test statistic. The second two use the same two test statistics and add an initial screening step to reduce the dimension of the data. In this paper we focus on the non-studentized method without screening. We have also implemented our proposed methods using the studentized test statistics and discovered it shows inferior performance in all experiments. The detailed results of the studentized test statistics are therefor only reported in the supplementary material.

To test for $H_{0}$ vs. $H_{1}$ in Equation (\ref{eq1}), we partition the $n$ observations into two sets, $\{ \bm{X}_{1}, ..., \bm{X}_{k} \}$ and $\{ \bm{X}_{k+1}, ..., \bm{X}_{n} \}$ at a split point $k \;(3\leq k\leq n-3)$. Each subset contains at least $3$ observations in order to calculate the sample means and variances. Therefore this methods can start monitoring with at least $6$ observations which is appropriate for monitoring short-run processes. As in \citet{chang2017simulation}, the non-studentized statistics ($NS$) with $n$ observations and split point $k$ are as follows, 
\begin{equation}\label{eq2}
T^{NS}_{n,k}=\text{max}_{1\leq r \leq p}\tfrac{\sqrt{k(n-k)}|\bar{X}_{k,r}-\bar{X}_{n-k,r}|}{\sqrt{n}}, 
\end{equation}
where $\bar{X}_{k,r}$ and $\bar{X}_{n-k,r}$ are the means of the $r$th variable ($X_{r}$) in the pre-shift sample and post-shift sample respectively. The monitoring statistic at time point $n$, referred as $U^{NS}_{n}$, is the maximum value of $T^{NS}_{n,k}$ over all split points $k$, 
\begin{equation}\label{eq3}
U^{NS}_{n}= \text{max}_{3\leq k \leq n-3}T^{NS}_{n,k}.
\end{equation}
The change-point $\tau$ can be estimated directly by the corresponding $k$ when $U^{NS}_{n}$ exceeds the control limits: 
\begin{equation}\label{eq4}
\hat{\tau}^{NS}_{n} = \text{arg}_{k}\,\text{max}_{3\leq k\leq n-3}(T^{NS}_{n,k}).
\end{equation}

Though \citet{chang2017simulation} present mathematical expressions for the critical values, they still use a fully data driven bootstrap method to estimate them. Therefore, we also use the same strategy to determine the control limits based on a predefined \emph{False Alarm Probability (FAP)}. 

\subsection{Moving Window}\label{3.2}
Note that for progressive monitors our statistic will be based on larger and larger samples (i.e. $n$ increases with time). The efficiency of detecting sparse shifts decreases with the sample size ($n$), because the signals are more likely to be buried by the noises. Another practical problem of our method is that the time and complexity of data processing multiply with incoming observations, because $n-5$ iterations are needed to compute the control statistic from $n$ observations. A third issue is that the statistic is influenced by the sample size, and therefore, the control limits are dynamic. To reduce the detection delay and promote the sensitivity of the proposed monitoring scheme, we add a moving window to the proposed method. 

For a fixed window size $W$, the monitoring procedure starts with the first $W$ observations $(\bm{X}_{1}, .., \bm{X}_{W})$. After collecting $s$ new observations, $s$ is the step size, it excludes the oldest $s$ observations and moves to the second window $(\bm{X}_{1+s}, .., \bm{X}_{W+s})$. An appropriate step size can speed up the computation and reduce the serial correlation, making the monitoring statistics become approximately independent. The observation matrix at time point $n\; (n>W)$ is $\mathbb{X}_{n}=(\bm{X}_{n-W+1}, \bm{X}_{n-W+2},...,\bm{X}_{n})$. The window-based non-studentized statistics are
\begin{equation}\label{eq5}
T^{NS}_{n,W,k^{*}}=\text{max}_{1\leq r \leq p}\tfrac{\sqrt{k^{*}(W-k^{*})}|\bar{X}_{k^{*},r}-\bar{X}_{W-k^{*},r}|}{\sqrt{W}}, 
\end{equation}
where $k^{*}\; (3 \leq k^{*}\leq W-3)$ is the split point inside the current window, and $\bar{X}_{k^{*},r}$ is the mean of first $k^{*}$ observations in window $W_{n}$. The corresponding charting statistic based on this window is 
\begin{equation}\label{eq6}
U^{NS}_{n, W}= \text{max}_{3\leq k^{*} \leq W-3}T^{NS}_{n,W, k^{*}}.
\end{equation}

Note that we observe $U^{NS}_{n, W}$ each $s$ time only. Estimating the change-point is also straightforward, if the control chart signals at time point $n$, the corresponding change-point estimate is 
\begin{equation}\label{eq7}
\hat{\tau}^{NS}_{n, W} = n-W+\text{arg}_{k^{*}}\,\text{max}_{3\leq k^{*} \leq W-3}T^{NS}_{n,W, k^{*}} .
\end{equation}
The statistic depends on the window size, which is a pre-specified fixed value. Consequently, the control limits are a constant value and they need to be simulated for different dimensionality and window sizes. Our method, referred by $NS_{W}$, signals when $U_{n,W}^{NS}>h_{p,W}$, where $h_{p,W}$ is the control limit. 

\subsection{Control Limits}\label{3.3}
In this paper, the control limits are determined by prespecifying the $FAP$ instead of the $ARL$, because the $ARL$ is affected by the window size and autocorrelation. Moreover, simulation for $ARL$s is more time consuming. We define the $FAP$ as the probability of at lease one false alarm in the process in time $1$ to $n$ \citep{chakraborti2008phase}. We use a bootstrap method to get the control limits with a predefined $FAP=\alpha$, as defined in Algorithm 1.

\begin{algorithm}
\SetAlgoLined
Step 1. Import $B$, $n$, $W$, $s$, and $\mathbb{X}_{p}$. Where $B$ is the number of bootstrap samples, $n$ is the monitoring interval, $W$ is the window size and $s$ is the step size. 

Step 2. Draw $B$ bootstrap samples of size $W$ with replacesment from $\mathbb{X}_{p}$. 

Step 3. Use Equation \ref{eq6} to calculate the charting statistics $U^{NS, b}_{W, W}$, $\; b=1, 2, ....B$ for every bootstrap sample. 

Step 4. Find the control limits as the $(1-\alpha)^{Q}$ empirical quantile from the pooled $U^{NS, b}_{W, W}$, $\; b=1, 2, ....B$ respectively with ${Q = \frac{1}{{max\{z \in \mathbb{Z} | z \leq (\frac{n-W}{s})\}+1}}}$. 
\caption{Bootstrap method for control limits}
\end{algorithm}

\begin{figure}[H] 
\includegraphics[width=\linewidth]{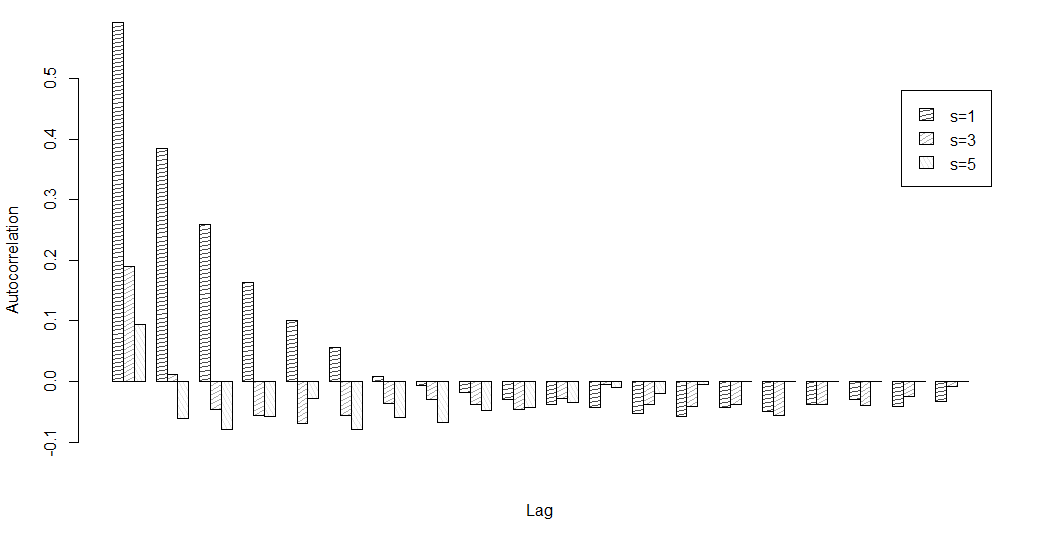}
\captionsetup{justification=centering}
\caption{The autocorrelation of $U_{n, W}^{NS}$ with various $s$, when $W=20$.}
\label{acf}
\end{figure}

Step 4 in Algorithm 1 is valid because appropriate step sizes can make the sequential $U_{n, W}^{NS}$ series an approximately independent process. Figure \ref{acf} shows the autocorrelation of $U_{n, W}^{NS}$ with $s=1, \,3, \, 5$, and $W=20$. When $s=5$, the autocorrelation is small for all lags. 

\begin{center}[Figure 1 near here.]\end{center}

In this paper, we set $\alpha= 0.01$, $B=10000$, $n=100$, $W=20, 30, 40$, $s=5$, and $\mathbb{X}_{p}$ is drawn from a $p$-dimensional normal distribution. All codes for calculating the statistics, control limits and performances analysis are available on GitHub (\url{https://github.com/wyfwzz/Supplementary-code-for-A-change-point-based-control-chart...-}). 

\section{Performance Study}\label{section 4}
\subsection{Experiments}\label{section 4.1}
To evaluate the OC performances of the proposed $NS_{W}$ method, we use simulation experiments. In these experiments, we vary the underlying distribution, dimensionality, window size, shift size, sparsity level and change-point. The detailed settings are as follow: 
\begin{enumerate}
\item  Models: the IC process follows a $p$-dimension standardized normal distribution $N_{p}(\bm{0}, \bm{I}_{p})$. To test the robustness to heteroscedasticity, correlation, and nonnormality, we design four OC models, where for each model  $\{ \bm{X}_i\}^{\tau}_{i=1} \sim N_{p}(\bm{0}, \bm{I}_{p})$ and 
\begin{itemize}
\item Model I: $\{ \bm{X}_i\}^{n}_{i=\tau+1} \sim N_{p}(\bm{\mu}_{1}, \bm{I}_{p})$. This model is the baseline model which only involves a change in the mean vector;
\item Model II: $\{ \bm{X}_i\}^{n}_{i=\tau+1} \sim N_{p}(\bm{\mu}_{1}, \lambda_{t}\bm{I})$, where $\lambda_{t}$ controls the change of variance and is set equal to $\{ 0.5, 0.6, ...., 0.9, 1, 0.9, ..., 0.6\}$ which is repeated until $n$. This model is used to simulate heteroscedasticity, after $\tau$, the variances vary with time;
\item Model III: $\{ \bm{X}_i\}^{n}_{i=\tau+1} \sim N_{p}(\bm{\mu}_{1}, \bm{\Sigma}_{1})$, where $ \bm{\Sigma}_{1} = (\sigma_{l,m})_{({1\leq l, m\leq p})}$, and $\sigma_{l,m} = 0.995^{|l-m|}$. This model incorporates the dependency among variables; 
\item Model IV: $\{ \bm{X}_i\}^{n}_{i=\tau+1} \sim t_{p, 30}(\bm{\mu}_{1}, \bm{\Sigma}_{1})$, where $ \bm{\Sigma}_{1} = (\sigma_{l,m})_{(1\leq l, m\leq p)}$, and $\sigma_{l,m} = 0.995^{|l-m|}$. This model is designed for testing the robustness to nonnormality;  
\end{itemize}
\item Dimensionality: $p=20,\, 50,\, 100$;
\item Window size: $W= 20,\, 30,\, 40$;
\item Change point: $\tau = 10,\,  25,\,  50$; 
\item Sparsity level (the percentage of OC variables): $v= 10\%,\, 25\%$, which means the first $v \times p$ variables change simultaneously after $\tau$;  
\item Shifts size: $\delta= 0.5,\, 1,\, 1.5,\, 2$, so that $\bm{\mu}_{1} = (\bm{\delta}_{vp}, \bm{0}_{p-vp})$, where $\bm{\delta}_{vp}$ is a $v \times p$ size vector with values of $\delta$ and $\bm{0}_{p-vp}$ is a $p-v \times p$ size vector with values of $0$. 
\end{enumerate}

\subsection{Performance Metrics}\label{section 4.2}
We run $R=1000$ simulation runs for every scenario to test the performance of the proposed method. In every simulation run, the monitoring procedure will stop once a signal is detected or when we run out of the predefined process with $n=100$ observations. The proportion of runs that are stopped by a signal to all runs represents the detecting power of the proposed method. This metric is called the \emph{Detection Rate (DR)}, where $DR=\frac{\sum_{j=1}^{R}{I(U_{n,W}^{NS, j} >h_{p,W}| H_{1}) }}{R}$. When $DR$ is close to $1$, it indicates that the proposed method is sensitive to change.

The $DR$ reflects the power of our method, in addition, a timely detection is also important. We define the \emph{Conditionally Expected Detection Delay (CED)} which is the time between the average stop point and the change point: $CED=\frac{\sum_{j=1}^{R}{min(arg_{n} (U^{NS, j}_{n,W}>h_{p,W} |H_{1}))}}{R}-\tau$. $CED$ is influenced by $W$, $s$, and $\tau$. When $W \leq \tau$, the smaller $CED$ reflects better performances. When $W > \tau$ and $CED = W-\tau$, it means that the control chart can detect a signal as early as the first window. All simulation results for all experiments can be found in the supplementary materials, along with the simulation results of the studentized method. We discuss the highlights in the rest of Section \ref{section 4}. 

\subsection{Baseline Simulation Results}\label{section 4.3}
Table \ref{baseline} shows the $DR$ and $CED$ for the $NS_{W}$ control chart under Model I with various $p$, $W$, $\delta$, $\tau$, and $v$. The $NS_{W}$ method has very low $DR$ when the shift is smaller than $1.5$. The $DR$ increases with the growth of window size. It also performs well in more sparse scenarios where $v=10\%$ with large change and appropriate window size. The location of $\tau$ has no significant influence on $DR$ unless it is very small ($\tau=10$). It has better $DR$ in detecting larger changes of higher dimensional mean vector. 

\begin{center}[Table 1 near here.]\end{center}

The $CED$ of the $NS_{W}$ method is worth considering when $DR \geq 0.5$. When $\tau>W$, the proposed method can detect a signal with about $10$ extra observations after the change point. Large $W$ can result in slower detection (large $CED$) especially if $\tau$ is very small. This result confirms the conclusion that the proposed method works well when the number of pre-shift and after-shift observations are balanced within one window. Therefore the optimal window size need to be determined based on a trade-off between sensitivity and detection delay. Consequently, we recommend the $NS_{W}$ method for detecting large and sparse shifts in high-dimensional mean vector with larger window size, and we select $W=40$ for the remaining experiments.  

\subsection{Robustness}\label{section 4.4}

The baseline model only considers mean shifts in multivariate normal distributed data, which is often an invalid assumption in practice. This section focuses on the results from Model II to IV to explore the robustness of our proposed method under heteroscedasticity, correlation, and nonnormality. The performance of our method is shown in Table \ref{robust}, for more results, see the supplementary material.

Compared with the baseline model, the $DR$ approximately decrease by $10$ percent points in heteroscedastic model with small shift size, and $DR=1$ when $\delta=2$. This result is consistent under different levels of sparsity. The $CED$ results show that heteroscedasticity has neglectable influence on the detection delay, and the $NS_{W}$ method is robust to data with unequal variances. The results from the other two models yield the same conclusion regarding of robustness. Therefore, we highly recommend to monitor a process with the $NS_{W}$ method when the underlying distribution is unknown or shows deviation from a normal distribution. 

\begin{center}[Table 2 near here.]\end{center}

\section{Diagnostic}\label{section 5}
After a signal is obtained, it is of interest to diagnose the problem. Equations (\ref{eq4}) and (\ref{eq7}) give the change-point estimate which can be useful in post-signal diagnosis. More than that, locating the assignable causes, the suspicious variables, is worth considering. One limitation of the proposed supremum-based method is that the signal is directly caused by one variable, and the abnormal behavior in other variables is ignored. We propose the following diagnostic procedure to select all suspicious variables: 
\begin{equation}\label{eq11}
\hat{V}^{NS}_{n} = \text{arg}_{r}(\tfrac{\sqrt{k^{*}(W-k^{*})}|\bar{X}_{k^{*},r}-\bar{X}_{W-k^{*},r}|}{\sqrt{W}}>h_{p,W}|U^{NS}_{n, W}>h_{p,W}, H_{1}),
\end{equation}
where $\hat{V}^{NS}_{n}$ is the set of variables that all yield a signal. To evaluate the accuracy of the diagnostic procedure,  we compared $\hat{V}^{NS}_{n}$ with the predefined variable set ${V}^{NS}_{n} = ({X}_{1}, .., X_{vp})$. The \emph{Detection Rate of Variables (DRV)} is defined as 
\begin{equation}\label{eq12}
DRV=P (\frac{|\hat{V}^{NS}_{n}\cap{V}^{NS}_{n}|}{|{V}^{NS}_{n}|}) = \frac{\sum_{i\in \hat{V}^{NS}_{n}}{I(i\leq p\times v)}}{p\times v}, 
\end{equation}
where $i$ is an element in $\hat{V}^{NS}_{n}$ (and the index of the variable that signals). 

Table \ref{diagnosis} shows the post-signal diagnostic performances of the proposed method with $p=100, \tau=25$, and $W= 40$ under different models. The \emph{Conditional Change-Point Estimate (CPE)} in Equation \ref{eq7} and $DRV$ are calculated from $1000$ simulations. The $NS_{W}$ method can estimate the change-point accurately under various setting of shift size and sparsity. There is no significant different performance between different models. It also shows good performance in finding the suspicious variables, especially when $\delta=2$, and is able to detect approximately $80\%$ of the change variables correctly. One possible way to improve the capability of locating assignable causes is adding a variable selection algorithm to the proposed method.  

\begin{center}[Table 3 near here.]\end{center}

\section{Comparison} \label{section 6}
As discussed in Section \ref{section 2}, \citet{chen2016distribution} proposed a distribution-free EWMA control chart (\emph{DFEWMA}). The \emph{DFEWMA} method is also based on a change-point model and is robust to nonnormality. It is efficient in detecting small and moderate shifts in location parameters for unknown distributions. 
We will compare the \emph{DFEWMA} control chart with our proposed method, because it is one of the most comparable methods that we can find and for the \emph{DFEWMA} method there is the matlab code available online. Before introducing the \emph{DFEWMA} chart, we first need to rewrite $H_{1}$ in Equation \ref{eq1} as 
\begin{equation*}
\bm{X}_{i} \sim
\begin{cases} 
N_{p}(\bm{\mu}_{0}, \bm{\Sigma}_{0}), & i = -m_{0}+1,..., 0, 1, ..., \tau, \\
N_{p}(\bm{\mu}_{1}, \bm{\Sigma}_{1}), & i>\tau \end{cases}. 
\end{equation*}
The difference is including $m_{0}$ IC observations to compose a reference sample. After collecting $\bm{X}_{n}$, the charting statistic can be constructed as $T_{n}(W, \lambda) = \sum_{r=1}^{p}{T^{2}_{n, r}(W, \lambda)}$, where
\begin{equation}\label{eq10}
T_{n, r}(W, \lambda) = \sum_{i=n-W+1}^{n}{(1- \lambda )^{n-i}}\frac {R_{n, i, r}-W(m_{0}+n+1)/2}{\sqrt{W(m_{0}+n+1)(m_{0}+n-W)/12}},
\end{equation}
where $W$ is the window size, $\lambda$ is the smoothing parameter, $R_{n, i, r}$ is the rank of $X_{i, r}$ among the sample ${X_{-m_{0}+1, r}, ..., X_{n, r}}$. The data-dependent control limits are determined online rather than before monitoring. For more detailed properties of this method, see \citet{chen2016distribution}. 

To make the \emph{DFEWMA} method more comparable with the proposed $NS_{W}$ methods, we use Model I and the same setting of parameters ($p, \tau, W$, and $v$) as in Section \ref{section 4}, and the shift size are set at $\delta = 1, 1.5, 2$. Next, $100$ IC observations are used as reference sample for the \emph{DFEWMA} method. Table \ref{compare} shows the simulation results of the $NS_{W}$ and \emph{DFEWMA} methods, when $\tau=50$ and $v= 10\%$, additional results can be found in the supplementary material. Both methods are designed to have a $FAP= 0.01$ with $100$ observations. The empirical $FAP$ are in brackets. \emph{DFEWMA} has larger empirical $FAP$ than the target value of $0.01$. We can neither fix the problem nor change the comparer, since the codes of other methods are unavailable.  

The $DR$ of the \emph{DFEWMA} chart is close to $1$ in all scenarios which outperforms the $NS_{W}$ method. We don't know how much it is affected by the larger $FAP$. It has better performance with less sparsity (large $v$), because it is a spatial-rank based statistic. The same explanation is also applicable to interpret the increase of $DR$ with dimensionality. Because the \emph{DFEWMA} chart uses increasing window size and moves with every new observation instead of using fixed window size and steps as us, it can achieve smaller detection delay. The proposed $NS_{W}$ method only outperforms in $CED$ with $W=20$ in lower dimension scenarios. One more reason for the good performance of the \emph{DFEWMA} method is that it has $100$ reference observations. As already pointed out by \citet{chen2016distribution}, the efficiency of the \emph{DFEWMA} chart relies on the large enough reference sample, which is unnecessary for our proposed method. An additional drawback of \citet{chen2016distribution} is the excessive false alarm rate. If there is a large reference sample and the practitioner does not mind many false alarms we recommend the \emph{DFEWMA} method. For all other scenarios we believe the practitioner should rather use our proposed method. It's detection power is a little smaller but monitoring can start quickly and false alarm is at the nominal level. 

\begin{center}[Table 4 near here.]\end{center}

\section{Case Study}\label{section 7}
In this section, we illustrate the proposed method by applying it to a real dataset from a semiconductor manufacturing process which is under constant surveillance via the monitoring of signals/variables collected from sensors and or measurement points. This dataset is available online (\url{http://archive.ics.uci.edu/ml/datasets/SECOM}). It consists of $590$ variables each with $1567$ records in chronological order. A classification label ($\pm1$) is given to indicate whether the product passes or fails, where $-1$ represents pass and $1$ is a failed one. Among these, $1463$ observations are the IC sample, and the remaining $104$ observations compose the OC sample. 

\begin{figure}[H] 
\includegraphics[width=\linewidth]{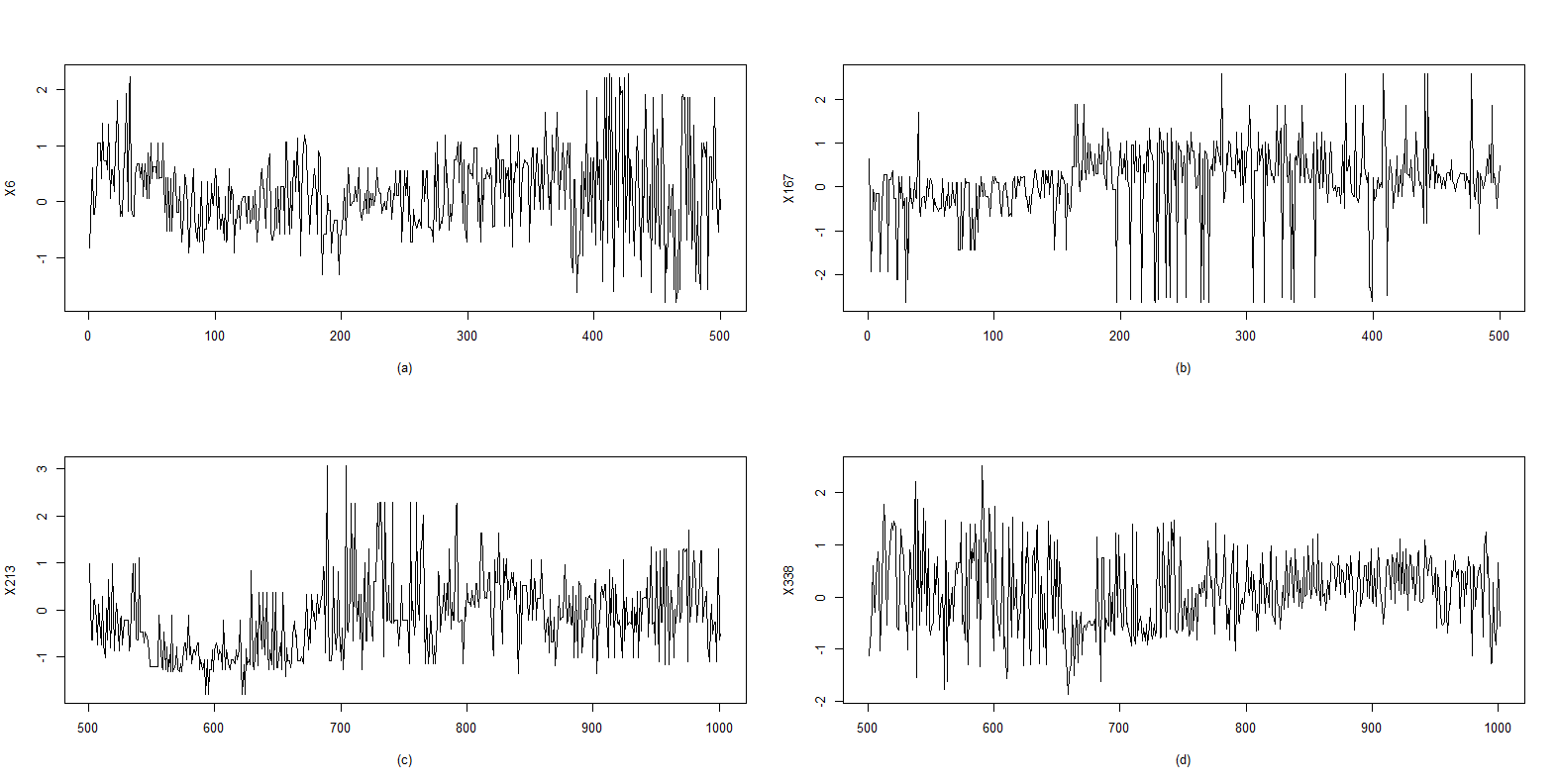}
\captionsetup{justification=centering}
\caption{The heteroscedasticity of standard scores of $X_{6}, \, X_{167}, \,X_{213}$, and $X_{338}$.}
\label{hete}
\end{figure}

Since the dataset contains constant values, null values and potential outliers, pre-processing is necessary. We remove the constant variables from both samples, leaving $416$ variables. The outliers in the IC sample are identified by Tukey's fences \citep{tukey1977exploratory}, and replaced by the the variable median. The missing values in both samples are replaced by variable median. After that, all the observations are standardized by the sample mean and standard deviation from the IC observations. The proposed method is applied to monitor the standard scores. Figure \ref{hete} shows the heteroscedastic standard scores of some variables when they are in control. 

\begin{center}[Figure 2 near here]\end{center}
The control limits are simulated by the proposed bootstrap method in Algorithm 1 based on the IC sample with $FAP=0.01$. Based on the assumption that the process starts from an IC status and then the mean vector changes at time $\tau$, where $\tau=10, 25, 50$. As shown in Table \ref{case}, the $NS_{W}$ method is efficient in monitoring real high-dimensional process. The DR is equal to $1$ in all scenarios. When $W>\tau$, it can always detect an out of control signal in the first window, and the $CPE$ is close to the real one. When $W<\tau$, $5$ observations detection delay is acceptable. These results confirm that the $NS_{W}$ method is robust to unknown distributions and heteroscedasticity. 

\begin{center}[Table 5 near here.]\end{center}

\section{Conclusion}\label{section 8}
Curse-of-dimensionality and unknown distributions increase the unreliability of  parameter estimate, especially when a large amount of IC observations are unavailable. In analyzing OC performance, sparsity is another practical assumption, which challenges both signal detection and diagnosis methods. In this research we proposed a change-point based control chart for monitoring sparse changes in high-dimensional mean vector, specifically for HDLSS scenarios. A moving window is added to speed up the computation and increase the sensitivity. As shown by the experimentation results, the proposed $NS_{W}$ chart is efficient in detecting large sparse shifts. And it can achieve accurate estimation of the change-point and the potential OC variables. It is robust to correlation, nonnormality, and heteroscedasticity, the latter is an important and often overlooked characteristic of high-dimensional process data. More than that, the experiments give some reference in determining the optimal window size, a tradeoff between sensitivity and false alarms needs to be considered. The real case study illustrates the robustness and practicability of the $NS_{W}$ method in application. Our proposed supremum-based method, the $NS_{W}$ method, is not as sensitive as the spatial-rank-based \emph{DFEWMA} chart in comparison. However the \emph{DFEWMA} needs a large reference sample and shows excessive false alarms, both are not issues in our method. Future directions for research is to improve the sensitivity. One possible way is to adapt a variable selection algorithms before starting monitoring. 

\section*{Supplementary Materials}
The PDF file provides the studentized method and the additional tables of the detailed simulation results. 

\section*{Acknowledgement}
The work of Inez M. Zwetsloot described in this paper was supported by a grant from the Research Grants Council of the Hong Kong Special Administrative Region, China (Project No. CityU 21215319). 

\section*{Declaration of Interest Statement}
Zezhong WANG is a Ph.D candidate in the City University of Hong Kong in the Department of Systems Engineering and Engineering Management. Her research interests include Statistical Process Monitoring, in particular, the High-Dimensional Process Monitoring. 

 \newpage
 \bibliographystyle{apalike}
 \bibliography{reference.bib}

\newpage
\begin{landscape}
\begin{table}[H]
\begin{center}
\captionsetup{justification=centering}
\caption{$DR$ and $CED$ of the $ NS_{W}$ method under Model I with various setting of $p$, $\delta$, $W$, $\tau$, and $v$.}
\begin{tabular}{ m{1em} m{1em} m{1.5em} m{3em} m{3em} m{3em} m{3em} m{3em} m{3em} m{1em} m{3em} m{3em}  m{3em} m{3em} m{3em} m{3em}} 
\hline &\\[-2ex]
& & & \multicolumn{6}{c}{$DR$} & & \multicolumn{6}{c}{$CED$} \\[0.2ex]
\cline{4-9} \cline{11-16} &\\[-2ex]
& & & \multicolumn{2}{c}{$\tau=10$} & \multicolumn{2}{c}{$\tau=25$} & \multicolumn{2}{c}{$\tau=50$} && \multicolumn{2}{c}{$\tau=10$} & \multicolumn{2}{c}{$\tau=25$} & \multicolumn{2}{c}{$\tau=50$}\\[0.2ex]
$p$ & $W$ & $\delta$ & $10\%$ &  $25\%$ & $10\%$ &  $25\%$ & $10\%$ &  $25\%$ && $10\%$ &  $25\%$ & $10\%$ &  $25\%$ & $10\%$ &  $25\%$\\[0.2ex]
\hline &\\[-1.5ex]
\multirow{6}{2em}{20} & \multirow{3}{2em}{20} & 1 & 0.078 & 0.114 & 0.063 & 0.119 & 0.073 & 0.13 && 23.8 & 17.7 & 13.8 & 11.1  & 9.2	& 9.9\\
& & 1.5 & 0.326 & 0.587 & 0.322 & 0.652 & 0.347 & 0.636 && 12.5 & 11.1 & 9.2 & 8.6 & 8.1 & 8.1\\
& & 2 & 0.772 & 0.984 & 0.82 & 0.993 & 0.831 & 0.979 && 10.5 & 10.1 & 7.3 & 6.0 & 7.3 & 5.8\\[1ex]
& \multirow{3}{2em}{40} & 1	& 0.074 & 0.167 & 0.236 & 0.495 & 0.273 & 0.514 && 32.4 & 32.4 & 19.0 & 18.5 & 17.5 & 16.4\\
& & 1.5 & 0.523 & 0.841 & 0.892 & 0.996 & 0.902 & 0.996 && 30.3 & 30.0 & 16.3 & 15.2 & 12.4 & 9.7\\
& & 2 & 0.964 & 1 & 1 & 1 & 1 & 1 && 30.0 & 30.0 & 15.0 & 15.0 & 7.6 & 5.9\\[0.5ex]
\hline &\\[-1.5ex]
\multirow{6}{2em}{50} & \multirow{3}{2em}{20} & 1 & 0.076 & 0.157 & 0.098 & 0.178 & 0.09 & 0.166 && 22.4 & 13.2 & 13.8 & 12.0 & 9.3 & 10.0\\
& & 1.5 & 0.467 & 0.791 & 0.537 & 0.825 & 0.506 & 0.807 && 11.2 & 10.3 & 9.0 & 8.0 & 8.5 & 7.5\\
& & 2 & 0.961 & 0.999 & 0.96 & 1 & 0.962 & 0.999 && 10.2 & 10.0 & 6.4 & 5.3 & 6.4 & 5.2\\[1ex]
& \multirow{3}{2em}{40} & 1 & 0.111 & 0.238 & 0.364 & 0.684 & 0.385 & 0.705 && 31.6 & 31.6 & 19.5 & 18.2 & 16.9 & 15.3\\
& & 1.5 & 0.751 & 0.968 & 0.987 & 1 & 0.992 & 1 && 30.1 & 30.0 & 15.4 & 15.0 & 10.5 & 8.2\\
& & 2 & 1 & 1 & 1 & 1 & 1 & 1 && 30.0 & 30.0 & 15.0 & 15.0 & 6.2 & 5.2\\[0.5ex]
\hline &\\[-1.5ex]
\multirow{6}{2em}{100} & \multirow{3}{2em}{20} & 1 & 0.031 & 0.112 & 0.042 & 0.105 & 0.052 & 0.114 && 16.5 & 11.7	& 11.3 & 9.5 & 11.6 & 9.3\\
& & 1.5 & 0.42 & 0.738 & 0.439 & 0.782 & 0.431 & 0.781 && 10.6 & 10.4 & 8.8 & 8.2 & 8.8 & 8.3\\
& & 2 & 0.973 & 1 & 0.982 & 1 & 0.977 & 1 && 10.1 & 10.0 & 6.5 & 5.3 & 6.5 & 5.2\\[1ex]
& \multirow{3}{2em}{40}	& 1 & 0.165 & 0.343 & 0.508 & 0.822 & 0.531 & 0.848 && 31.6 & 30.6 & 18.7 & 17.3 & 16.3 & 14.9\\
& & 1.5 & 0.903 & 0.996 & 1 & 1 & 1 & 1 && 30.1 & 30.0 & 15.1 & 15.0 & 9.1 & 6.9\\
& & 2 & 1 & 1 & 1 & 1 & 1 & 1 && 30.0 & 30.0 & 15.0 & 15.0 & 5.4 & 5.0\\[0.5ex]
\hline
\end{tabular}\label{baseline}
\end{center}
\end{table}
\end{landscape}

\begin{table}[H]
\begin{center}
\captionsetup{justification=centering}
\caption{$DR$ and $CED$ for the proposed $NS_{W}$ method under Model I to IV with various setting of $\delta$ and $v$. When $p=100$, $\tau=25$, and $W=40$.}
\begin{tabular}{ m{8em} m{2em} m{4em} m{4em} m{0.2em}  m{4em}  m{4em} } 
\hline &\\[-2ex]
& & \multicolumn{2}{c}{$DR$} & & \multicolumn{2}{c}{$CED$} \\[0.1ex]
\cline{3-4}   \cline{6-7} &\\[-2ex]
$Model$ & $\delta$ & $v=10\%$ & $v=25\%$ & & $v=10\%$ & $v=25\%$ \\[0.1ex]
\hline &\\[-2ex]
\multirow{3}{2em}{I (Baseline)} & 1 & 0.508 & 0.822 & & 18.7 & 17.3\\
& 1.5 & 1 & 1 & & 15.1 & 15.0 \\
& 2 & 1 & 1 & & 15.0 & 15.0 \\[1ex]
\multirow{3}{2em}{II (Heteroscedastic)} & 1 & 0.408 & 0.712 & & 19.7 & 18.6 \\
& 1.5 & 0.999	& 1 & & 15.1 & 15.0 \\
& 2 & 1 & 1 & & 15.0 & 15.0\\[1ex]
\multirow{3}{2em}{III (Dependent)} & 1 & 0.348 & 0.539 & & 19.2 & 19.5 \\
& 1.5 & 0.968 & 0.992 & & 16.3 & 15.6 \\
& 2 & 1 & 1 & & 15.0 & 15.0\\[1ex]
\multirow{3}{2em}{IV (Nonnormal)}	& 1 & 0.404 & 0.547 & & 18.9 & 19.2 \\
& 1.5 & 0.955 & 0.989 & & 16.3 & 15.7 \\
& 2 & 1 & 1 & & 15.0 & 15.0 \\[0.5ex]
\hline
\end{tabular}\label{robust}
\end{center}
\end{table}

\begin{table}[H]
\begin{center}
\captionsetup{justification=centering}
\caption{Post-signal diagnosis of the $NS_{W}$ chart under Model I to Model IV with various setting of $v$. When $p=100$, $\tau=25$, $W=40$.}
\begin{tabular}{ m{8em} m{2em} m{4em}  m{4em} m{0.5em}   m{4em} m{4em} } 
\hline &\\[-2ex]
& & \multicolumn{2}{c}{$CPE$} & & \multicolumn{2}{c}{$DRV$}\\
\cline{3-4}   \cline{6-7} &\\[-2ex]
$Model$ & 	$\delta$ & $v=10\%$ & $v=25\%$ & & $v=10\%$ & $v=25\%$\\
\hline & \\[-2ex]
\multirow{3}{2em}{I (Baseline)} & 1 & 25.4 & 25.1 && 0.105 & 0.045 \\
& 1.5 & 24.9 & 24.9 && 0.261 & 0.263\\
& 2 & 24.9 & 24.9 && 0.805 & 0.814\\[1ex]
\multirow{3}{2em}{II (Heteroscedastic)} & 1 & 24.3 & 24.2 && 0.103 & 0.043\\
& 1.5 & 24.6 & 24.6 && 0.263 & 0.241\\
& 2 & 24.8 & 24.8 && 0.843 & 0.836\\[1ex]
\multirow{3}{2em}{III (Dependent)} & 1 & 25.2 & 24.7 && 0.179 & 0.090\\
& 1.5 & 24.9 & 24.7 && 0.405 & 0.338\\
& 2 & 24.8 & 24.8 && 0.828 & 0.807\\[1ex]
\multirow{3}{2em}{IV (Nonnormal)} & 1 & 24.8 & 24.8 && 0.200 & 0.121\\
& 1.5 & 24.7 & 24.7 && 0.396 & 0.350\\
& 2 & 24.8 & 24.8 && 0.818 & 0.821\\[0.1ex]
\hline
\end{tabular}\label{diagnosis}
\end{center}
\end{table}

\begin{table}[H]
\begin{center}
\captionsetup{justification=centering}
\caption{The $DR$ and $CED$ of the \emph{DFEWMA} and $NS_{W}$ charts with various combinations of $p$, $W$, and $\delta$. When $m_{0}=100$, $v=10\%$, $\tau=50$. Between bracket the $FAP=0.01$.}
\begin{tabular}{ m{2em} m{2em} m{3em} m{4.5em} m{4.5em} m{0.5em} m{4.5em} m{4.5em}} 
\hline &\\[-2ex]
& & & \multicolumn{2}{c}{$DR$} & & \multicolumn{2}{c}{$CED$}\\[0.1ex]
\cline{4-5}   \cline{7-8} \\[-2ex]
$p$ & $W$ & $\delta$ & $NS_{W}$ & $\emph{DFEWMA}$ & & $NS_{W}$ & $\emph{DFEWMA}$ \\[0.1ex]
\hline &\\[-2ex]
\multirow{9}{2em}{20} & \multirow{3}{2em}{20} & 0 & (0.015) & (0.05) & & & \\	
& & 1.5 & 0.347 & 1 & & 8.08 & 11.01\\
& & 2 & 0.831 & 1 & & 7.30 & 8.07\\[0.5ex]
& \multirow{3}{2em}{30} & 0 & (0.013) & (0.042)	 & & & \\
& & 1.5 & 0.715 & 1 & & 11.21 & 11.20\\
& & 2 & 0.991 & 1 & & 7.51 & 8.23\\[0.5ex]
& \multirow{3}{2em}{40}& 0 & (0.01) & (0.046) & & &\\	
& & 1.5 & 0.902 & 1 & & 12.39 & 10.51\\
& & 2 & 1 & 1 & & 7.64 & 8.03 \\[0.1ex]
\hline &\\[-2ex]		
\multirow{9}{2em}{50} & \multirow{3}{2em}{20} & 0 & (0.018) & (0.047) & & & \\	
& & 1.5 & 0.506 & 1 & & 8.54 &	7.19\\
& & 2 & 0.962 & 1	& & 6.36 &	5.40\\[0.5ex]
& \multirow{3}{2em}{30} & 0 & (0.013) &	(0.037) & & & \\
& & 1.5 & 0.918 & 1 & & 10.11 & 7.23\\
& & 2 & 1 & 1 & & 6.30 & 5.30\\[0.5ex]
& \multirow{3}{2em}{40}& 0 & (0.016) &	(0.043) & & &\\	
& & 1.5 &0.992 & 1 & & 10.50 & 7.14\\
& & 2 & 1 & 1 & & 6.21 & 5.64\\[0.1ex] 
\hline &\\[-2ex]
\multirow{9}{2em}{100} & \multirow{3}{2em}{20} & 0 & (0.004) & (0.047) & & & \\	
& & 1.5 & 0.431 & 1 & & 8.84 & 5.33\\
& & 2 & 0.977 & 1	& & 6.51 & 4.34\\[0.5ex]
& \multirow{3}{2em}{30} & 0 & (0.011) &	(0.047) & & & \\
& & 1.5 & 0.966 & 1 & & 9.65 & 5.33\\
& & 2 & 1 & 1 & & 5.61 & 4.21\\[0.5ex]
& \multirow{3}{2em}{40}& 0 & (0.003) &	(0.037) & & &\\	
& & 1.5 & 1 & 1 & & 9.11 & 5.54\\
& & 2 & 1 & 1 & & 5.39	& 4.45\\ [0.1ex]
\hline
\end{tabular}\label{compare}
\end{center}
\end{table}

\begin{table}[H]
\begin{center}
\captionsetup{justification=centering}
\caption{The $DR$ and $CED$ of the proposed $NS_{W}$ method in signal detection, and the $CPE$ in post-signal diagnosis with various setting of $W$ and $\tau$.}
\begin{tabular}{ m{2.5em} m{2.5em} m{2.5em} m{2.5em} m{0.1em} m{2.5em}  m{2.5em} m{2.5em} m{0.1em} m{2.5em} m{2.5em} m{2.5em}} 
\hline &\\[-2ex]
$W$ & \multicolumn{3}{c}{20} && \multicolumn{3}{c}{30} && \multicolumn{3}{c}{40}\\[0.1ex]
\cline{1-4} \cline{6-8} \cline{10-12} &\\[-2ex]
$\tau$ & 10 & 25 & 50 && 10 & 25 & 50 && 10 & 25 & 50\\
\hline &\\[-2ex]
$DR$ & 1 & 1 & 1 && 1 & 1 & 1 && 1 & 1 & 	1\\
$CED$ & 10.00 & 4.99 & 4.92 && 20.00 & 5.02 & 4.95 && 30.00 & 15.00 & 4.93 \\
$CPE$ & 14.15 & 26.24 & 51.19 && 20.05 & 26.35 & 51.22 && 25.96 & 31.94 & 51.12\\
\hline
\end{tabular}\label{case}
\end{center}
\end{table}

\begin{itemize}
  \item Figure 1. The autocorrelation of $U_{n, W}^{NS}$ with various $s$, when $W=20$.
  \item Figure 2. The heteroscedasticity of standard scores of $X_{6}, \, X_{167}, \,X_{213}$, and $X_{338}$.
\end{itemize}

\end{document}